\documentclass[twocolumn,floatfix,altaffilletter,superscriptaddress,preprintnumbers,
               tightenlines,showpacs,showkeys,nofootinbib]{revtex4-1}
\usepackage[utf8]{inputenc}
\usepackage[colorlinks=true,citecolor=blue,linkcolor=blue]{hyperref}
\usepackage[normalem]{ulem}
\usepackage{amsmath,amssymb, mathrsfs}
\usepackage{epsfig}
\usepackage{graphicx}               % Standard graphics package
\usepackage{url}
\usepackage{color}
\usepackage[dvipsnames]{xcolor}
\usepackage{soul}
\usepackage[capitalise, english]{cleveref}
\usepackage{siunitx}
\usepackage{xspace}
\usepackage{enumitem}
\usepackage{multirow}
\usepackage{scalerel}
\usepackage{bbding}
\usepackage{adjustbox}

\newcommand\myshade{80}
\colorlet{mylinkcolor}{ForestGreen}
\colorlet{mycitecolor}{Red}
\colorlet{myurlcolor}{violet}

\hypersetup{
  linkcolor  = mylinkcolor!\myshade!black,
  citecolor  = mycitecolor!\myshade!black,
  urlcolor   = myurlcolor!\myshade!black,
  colorlinks = true
}
\DeclareSIUnit\parsec{pc}
\DeclareMathOperator{\BR}{BR}

\newcommand{\lam}{{\ensuremath{\lambda}}}

 \definecolor{tobycolour}{rgb}{.5,.0,.5}

\newcommand{\gsim}{\stackrel{>}{\sim}}
 
 \definecolor{myred}{rgb}{0.886275, 0.290196, 0.2}
 \definecolor{mygreen}{rgb}{0.556863, 0.729412, 0.258824}
 \definecolor{myblue}{rgb}{0.203922, 0.541176, 0.741176}

\begin{document}

\newcommand{\AddrBonn}{%
Bethe Center for Theoretical Physics \& Physikalisches Institut der 
Universit\"at Bonn, Nu{\ss}allee 12, 
 53115 Bonn, Germany
}

\newcommand{\AddrCern}{%
Theoretical Physics Department, CERN,
             Esplanade des Particules, 1211 Geneva 23, Switzerland}
             
\newcommand{\AddrDesy}{%
 Deutsches Elektronen-Synchrotron DESY, Notkestr. 85, 22607 Hamburg, Germany
 }

\newcommand{\AddrUCB}{
Berkeley Center for Theoretical Physics, University of California, Berkeley, CA 94720
}

\newcommand{\AddrLBL}{
Theoretical Physics Group, Lawrence Berkeley National Laboratory, Berkeley, CA 94720
}
%=============================================================================
\title{Lepton PDFs and Multipurpose Single-Lepton Searches at the LHC}

\author{Herbi K. Dreiner}\email{dreiner@uni-bonn.de}
\affiliation{\AddrBonn}

\author{V\'ictor Mart\'in Lozano}\email{victor.lozano@desy.de}
\affiliation{\AddrDesy}

\author{Saurabh Nangia}\email{nangia@physik.uni-bonn.de}
\affiliation{\AddrBonn}

\author{Toby Opferkuch}\email{toby.opferkuch@cern.ch}
\affiliation{\AddrUCB}
\affiliation{\AddrLBL}
\affiliation{\AddrCern}

\preprint{BONN-TH-2021-13, DESY 21-210}
%=============================================================================

\begin{abstract}
\noindent
A final state consisting of one charged lepton, at least one jet, and little missing transverse 
energy can be a very promising signature of new physics at the LHC across a wide range of models. 
However, it has received only limited attention so far. In this work we discuss the potential 
sensitivity of this channel to various new physics scenarios. To demonstrate our point, we 
consider its application to lepton parton distribution functions (PDFs) at the LHC in the context of 
supersymmetry. These lepton PDFs can lead to resonant squark production (similar to leptoquarks) 
via lepton number violating couplings present in $R$-parity Violating Supersymmetry (RPV-SUSY). 
Unlike leptoquarks, in RPV-SUSY there are many possible decay modes leading to a wide range of 
signatures. We propose
two generic search regions: (a) A single first or second generation charged lepton, exactly 1~jet
and low missing transverse energy, and (b) A single first or second generation charged lepton, 
at least 3 jets, and low missing transverse energy. We demonstrate that together these cover a 
large range of RPV-SUSY signatures, and have the potential to perform better than existing 
low-energy bounds, while being general enough to extend to a wide range of possible models 
hitherto not explored at the LHC. 
\end{abstract}

\maketitle

%=============================================================================
\section{Introduction}
\label{sec:intro}
%=============================================================================

With the High-Luminosity (HL) era almost upon us, the Large Hadron Collider (LHC) is gearing 
up for a paradigm shift: A transition from energy upgrades to a focus on increased luminosity. 
The LHC has already accumulated close to $\SI{200}{\femto\barn^{-1}}$ of data \cite{lhc-lumi}. 
This will be surpassed in just a single year of HL-LHC runtime, which by its conclusion will
increase the total integrated luminosity by a factor $15$ overall \cite{Apollinari:2017lan}.

Given the breadth of possible observables, such a drastic increase in integrated luminosity places 
a renewed importance in exploring the question, ``Where should we look for new physics?''. In 
contrast to an increase in energy, increases in luminosity do not typically yield dramatic
improvements in reach when considering resonantly produced new physics. On the other hand, rare 
processes, indirect searches, and new trigger techniques -- to which the experiments are 
currently insensitive with the data on tape -- can offer promising avenues to explore; current examples include
Refs.~\cite{CMS-PAS-FTR-18-018,Bhattacherjee:2020nno,Gershtein:2017tsv,Gershtein:2019dhy,Evans:2020aqs,Gershtein:2020mwi}. Such strategies can be more powerful than one may expect. In this work we demonstrate this point explicitly by making the case for a specific example of an unusual signal 
at the LHC: One light charged lepton, one light jet, and no missing energy -- what we will call the \textit{single-lepton channel}.

We mention that an interesting aspect of the kind of lepton-number violating searches we discuss is the fact that they could also be relevant to current topics of interest such as the $B$-anomalies \cite{LHCb:2017rln,LHCb:2021trn}, and the muon $(g-2)$ \cite{Muong-2:2002wip, Muong-2:2004fok, Muong-2:2006rrc, Muong-2:2021ojo}; the variety of models that produce signals of interest to us may also explain these anomalies (see for example Refs.~\cite{Bauer:2015knc, FileviezPerez:2021lkq,Trifinopoulos:2019lyo, Hu:2020yvs, Chakraborti:2022vds}).

The outline of this paper is as follows. In \cref{sec:single-lepton-channel}, we discuss the 
single-lepton channel in detail, and argue that it can be applied to a wide range of models with lepton number violating interactions. \cref{sec:rpv-applic} discusses its application to the specific context of RPV-SUSY. In \cref{sec:implementation}, we demonstrate how such a search for RPV-SUSY may be implemented at the LHC.  In \cref{sec:discussion}, we present and discuss our numerical results. We conclude in \cref{sec:conclusions}.

%%=============================================================================
\section{The Single-Lepton Channel}
\label{sec:single-lepton-channel}
%%=============================================================================

The final state we are interested in has exactly one energetic\footnote{There can be additional soft 
objects from the showering but they will not affect the analysis.} first or second generation charged 
lepton $\left(\ell\right)$, at least one energetic light jet $\left(j\right)$, and little or no missing transverse energy ($E^{
\text{miss}}_\text{T}$). In what follows we refer to this as the single-lepton channel for short. 

At first glance this channel is forbidden at hadron colliders in the limit $E^\text{miss} _\text{T} 
\rightarrow 0$ as the final state is odd in lepton number (discounting the possibility of additional 
un-tagged soft leptons). Such a channel could still be populated if at least one of the following 
ingredients is present:
\begin{enumerate}
    \item Sizeable lepton number violating interactions.
    \item A hard process that is initiated by the leptonic content of the proton.
\end{enumerate}
However, the first is bounded by strict low-energy constraints; see, for instance,  
Refs.~\cite{Dreiner:1997uz,Barbier:2004ez}. For the second, the leptonic parton distribution
functions (PDFs) are suppressed compared to quark or gluon initiated processes as they rely 
on the splitting function of the photon. The photon density in the proton is low and the 
splitting adds one more power of the fine-structure constant, $\alpha_\text{EM}$. Early work on lepton PDFs can be found in Refs.~\cite{Ohnemus:1994xf, Bertone:2015lqa}. A higher order 
calculation has recently been performed in Ref.~\cite{Buonocore:2020nai}.

Before turning to the details of the proposed search, we first discuss existing single-lepton
searches in the literature. Among early experimental work, the only potentially sensitive 
searches are those by \texttt{CMS} and \texttt{ATLAS} for quantum black holes, such as 
Refs.~\cite{CMS:2013ndg,ATLAS:2014wsp}. However, these searches require very high multiplicity 
final states producing a large overall scalar sum $\sum p_T\gsim \SI{2}{\tera\electronvolt}$, 
have no upper limit on $E^{\text{miss}}_\text{T}$, and allow for more than 1 charged lepton. 
In Refs.~\cite{ATLAS:2017irs, CMS:2017yoc}, attempts were made at model-independent searches 
by considering several hundreds of signal topologies including the ones we are studying here. 
However, since the datasets considered correspond to small integrated luminosities, and the 
analyses are not designed to optimally target the single-lepton final state, we expect 
low sensitivity to our rare signals.

Another related channel was proposed in Ref.~\cite{Lisanti:2011tm} (and searched for at the LHC 
in Refs.~\cite{ATLAS:2017oes,CMS:2017szl,ATLAS:2021fbt,CMS:2021knz}) involving a single lepton 
with high jet multiplicity (1$\ell+nj$, $n$ large) but no $E^{\text{miss}}_\text{T}$ cut. This 
was constructed to be sensitive to several new physics scenarios which may escape high 
$E^{\text{miss}}_\text{T}$ searches. These include lepton number conserving models, in which 
case the possibly present $E^{\text{miss}}_\text{T}$ from neutrinos is diluted due the large 
number of final state objects. In later work, Ref.~\cite{Evans:2013jna} demonstrated how such a 
channel can play a pivotal role in closing the last remaining gaps in natural supersymmetric 
theories. But the channel is sensitive to more generic models of new physics as well, 
\textit{e.g.}, composite Higgs models, models producing top-rich final states, or even more 
exotic phenomena involving  high-scale non-perturbative effects.

We propose to go beyond this earlier work by focusing on the related but orthogonal final state: 
1$\ell+nj$, $n$ \textit{small}. The main difference is that we allow for a significantly lower 
multiplicity in the final state. The low $E^{\text{miss}}_\text{T}$ in such scenarios is not 
due to dilution as above, but due to lepton number violating processes and/or lepton PDFs. Thus, 
unlike above, we require a strict upper limit on $E^{\text{miss}}_\text{T}$. In 
\cref{sec:implementation}, we define two separate search regions to cover what we think are the 
most relevant scenarios missed so far. 

Very recently, Ref.~\cite{Buonocore:2020erb} demonstrated that a specific example of the above -- 
a final state with one energetic charged lepton, low $E^{\text{miss}}_\text
{T}$, and exactly one energetic jet -- may probe large unexplored regions of the leptoquark parameter space at 
the LHC.\footnote{Also see Ref.~\cite{Haisch:2020xjd} for a generalization to the case of third-generation leptons and jets.} The $s$-channel leptoquark resonance is produced via the leptonic PDFs of 
the proton, mimicking the production at HERA \cite{Buchmuller:1986zs,Butterworth:1992tc,Dreiner:1994tj}. While the result may seem surprising at first due to the suppressed lepton PDFs, the point is that the $s$-channel resonance has double the kinematic reach compared to leptoquark pair production, and smaller suppression from the
leptoquark coupling compared to the Drell-Yan mode. Thus, it can complement these modes~\cite{Crivellin:2021egp} by probing regimes where the leptoquark mass is beyond the pair production threshold, while the leptoquark coupling is not large enough for Drell-Yan to be effective. Further, it has a far cleaner signature and a dynamic boost compared to other single production 
modes. This idea forms the basis for the work that follows: How can these single-lepton searches be generalized to 
exploit a wide range of new physics appearing in $s$-channel resonances. 

Searches for 1$\ell+nj$ with $n$ small are yet to be
performed. These would not only target leptoquarks but more generally theories that contain lepton 
number violation, or even some new interaction between leptons and quarks, such as a heavy partially
leptophillic $Z'$~\cite{Buras:2021btx}. As an example of the former we will consider $R$-parity Violating SUSY in what follows.

%%=============================================================================
\section{An Application to R-parity Violating Supersymmetry}
\label{sec:rpv-applic}
%%=============================================================================
\begin{figure}
	\centering
\hspace{1cm}\includegraphics[width = 0.4\textwidth]{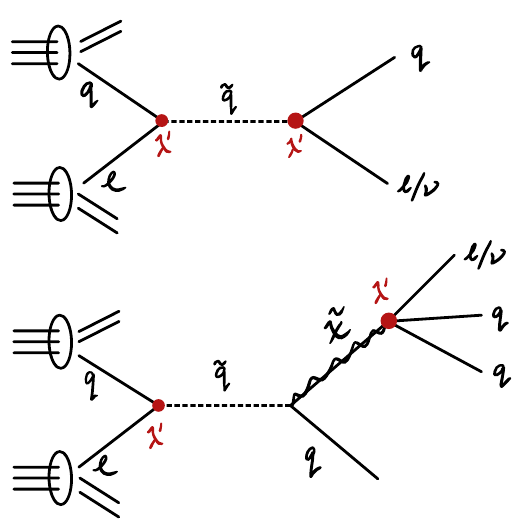}
	\caption{Resonant squark produced via the $\lambda'_{ijk}L_iQ_j\Bar{D}_k$ operator at the LHC followed by its direct decay mode (top) and decay via an example of a gauge-cascade mode (bottom). Here $\tilde\chi$ denotes a chargino or neutralino, lighter than the squark.}
	\label{fig:feyn_graphs}
\end{figure}

RPV is the most general realization of the minimal supersymmetric Standard Model (MSSM) where,
with the minimal field content, all renormalizable operators allowed under the Standard Model (SM) gauge symmetries
are permitted \cite{Weinberg:1981wj}. This has immediate phenomenological consequences, allowing
for lepton- and baryon number violating operators. However, a subset must be prohibited (for 
example, through a discrete symmetry) to ensure the stability of the proton 
\cite{Ibanez:1991pr,Dreiner:1997uz,Allanach:2003eb,Dreiner:2003yr,Dreiner:2006xw,Dreiner:2012ae,Csaki:2013jza}. 
In the MSSM, the imposed discrete symmetries \cite{Farrar:1978xj,Dreiner:2005rd} -- \textit{e.g.} $R$-parity -- prohibit the entire set of lepton- and baryon number violating operators.

Allowing some of the RPV terms changes the phenomenology compared to the MSSM in two drastic ways: (a) The 
lightest supersymmetric particle (LSP) is unstable, and (b) Single production of supersymmetric
particles is possible. The latter not only improves our kinematic reach but also provides a mechanism for overcoming the suppression from the lepton PDFs. 

The complete superpotential for the RPV-MSSM is given by,
 \begin{equation}
 W = W_{\mathrm{MSSM}} + W_{\mathrm{LNV}} + W_{\mathrm{BNV}}\,,
 \label{eq:eq1}
 \end{equation}
where $W_{\mathrm{MSSM}}$ is the usual MSSM superpotential -- for details see Ref.~\cite{Allanach:2003eb} -- and
\begin{align}
W_{\mathrm{LNV}} &= \frac{1}{2}\lambda_{ijk}L_iL_j\overline{E}_k + \lambda'_{ijk}L_iQ_j\overline{D}_k + 
\kappa_{i}H_uL_i\,, \label{eq:LNV}\\
W_{\mathrm{BNV}} &= \frac{1}{2}\lambda''_{ijk}\overline{U}_i\overline{D}_j\overline{D}_k\,,
\label{eq:BNV}
\end{align}
are the new interactions that explicitly violate $R$-parity. In the above, $L\,(Q)$ and $\Bar{E}\,(\Bar{U},\Bar{D})$ refer to the lepton (quark) $\mathrm{SU}(2)_{L}$ doublet and singlet chiral superfields from the MSSM, respectively, while $H_u, H_d$ label the $\mathrm{SU}(2)_{L}$ doublet Higgs chiral superfields. The $\lam$'s, are dimensionless coupling constants; the $\kappa$'s are dimension-one mass parameters. All gauge indices are suppressed but we explicitly write the generational ones: $i,j,k = 1, 2, 3$, with a summation implied over repeated labels.

\begin{table*}
\begin{center}
\begin{ruledtabular}
\renewcommand{\arraystretch}{1.2}
\begin{tabular}{llc} 
{\bf Cascade End} & {\bf Example Decay Chain} & {\bf Final State Signature}\\
Bino ($\widetilde{B}$) & $\Tilde{d}_R \rightarrow \Tilde{B} + 1j$ & $1\ell + 3j$\\
Wino ($\widetilde{W}$) & $\Tilde{d}_R \rightarrow \Tilde{g}^{\left(*\right)} + 1j \rightarrow \Tilde{q}^{\left(*\right)}_L + 2j \rightarrow \widetilde{W}^0/\widetilde{W}^{\pm} + 3j$ & $1\ell + 5j$\\
Gluino ($\Tilde{g})$ & $\Tilde{d}_R \rightarrow \Tilde{g} + 1j$ & $1\ell + 3j$\\
Doublet squark ($\Tilde{q}_L$) & $\Tilde{d}_R \rightarrow \Tilde{g}^{\left(*\right)} + 1j \rightarrow \Tilde{q}_L + 2j$ & $1\ell + 3j$\\
Up-type singlet squark ($\Tilde{u}_R$) & $\Tilde{d}_R \rightarrow \Tilde{g}^{\left(*\right)} + 1j \rightarrow \Tilde{u}_R + 2j$ & $1\ell + 5j$\\
Doublet charged slepton ($\Tilde{e}_L$) & $\Tilde{d}_R \rightarrow \Tilde{g}^{\left(*\right)} + 1j \rightarrow \Tilde{q}^{\left(*\right)}_L + 2j \rightarrow \widetilde{W}^{0\left(*\right)} + 3j \rightarrow \Tilde{e}_L + 1\ell + 3j$ & $1\ell + 5j$ \\
Sneutrino ($\Tilde{\nu}_L$) & $\Tilde{d}_R \rightarrow \Tilde{g}^{\left(*\right)} + 1j \rightarrow \Tilde{q}^{\left(*\right)}_L + 2j\rightarrow \widetilde{W}^{\pm\left(*\right)} + 3j \rightarrow \Tilde{\nu}_L + 1\ell + 3j$ & $1\ell + 5j$  \\
Singlet charged slepton ($\Tilde{e}_R$) & $\Tilde{d}_R \rightarrow \widetilde{B}^{\left(*\right)} + 1j \rightarrow \Tilde{e}_R + 1\ell + 1j$ & $3\ell + 3j$\\
\end{tabular}
\end{ruledtabular}
\end{center}
\caption{Decay chains and final state signatures resulting from a resonantly produced down-type squark 
($\tilde{d}_R$) in the case of $\lam' \neq 0$. The first column indicates the supersymmetric particle involved in the final step of the decay chain, which is typically the LSP. We give a representative decay chain for each case that 
populates the single-lepton channel, as well as the total signature in the final column. $\left(*\right)$ indicates possibly off-shell.}
\label{tab:tab1}
\end{table*}

Consider just a single LQD operator in the superpotential, \textit{e.g.}, $\lam'_{111}L_1Q_1\bar D_1$. The 
simplest possibility leading to a single-lepton signal via this operator requires a spectrum with only 
a light squark, $\tilde{d}_R$ or $\tilde{u}_L$. Given no other supersymmetric particles lighter than the 
squark, and taking into account the lepton PDFs, the dominant process is an $s$-channel squark resonance, 
illustrated in the top diagram of \cref{fig:feyn_graphs}. Here the squark decays back to the initial state, as it is the LSP. 
This is analogous to the scalar leptoquark scenarios considered in Ref.~\cite{Buonocore:2020erb}, 
leading to a final state comprised of a single lepton plus one jet. In contrast to leptoquark models,
supersymmetry typically predicts new states lying below the squark masses 
\cite{Allanach:1999mh,Allanach:2001kg}, such that the squark would cascade via gauge couplings, typically all the way down to the LSP. In \cref{tab:tab1} we list possible final state signatures for $\tilde{d}_R$ as a function of the particular state 
lying below the squark mass. The branching ratios into these final 
states is determined via the relative size of the RPV coupling versus the gauge couplings, as well as 
details of the mass spectrum, see Refs.~\cite{Dreiner:1991pe,Dreiner:2012wm,Dercks:2017lfq}. A canonical 
example is an LSP bino-like neutralino \cite{Dreiner:2009ic} giving rise to the extended decay 
chain depicted in the bottom diagram of \cref{fig:feyn_graphs}. Upon its production from the squark, the 
bino-like neutralino decays via the RPV operator resulting in a final signature with three jets plus the desired single lepton. 

Surveying the final states in the last column of \cref{tab:tab1}, we observe that the single-lepton channel 
can be populated irrespective of the supersymmetric particle involved in the final step of the decay 
cascade. The only exception is a decay chain featuring a light charged slepton $\tilde{e}_R$, where two
additional charged leptons result.\footnote{The case of additional leptons is more constrained by existing
searches, see Ref.~\cite{Evans:2013jna}.} There are however limitations to this analysis. As stated above 
the cascade details depend sensitively on the mass spectrum, as well as the size of the RPV coupling, $\lambda^\prime$, relative to the gauge couplings \cite{Dreiner:2012wm}. Fortunately, the latter does not modify the above conclusions. 
The large value of the strong coupling means that additional steps in the decay chains typically only increase 
jet multiplicity. Therefore, the single-lepton channel provides a sensitive probe irrespective of the 
model details, allowing us to implement a largely model-independent search strategy. We now turn to the 
details of how such a strategy can be implemented at the LHC.

%=============================================================================
\section{Implementation}
\label{sec:implementation}
%%=============================================================================
We first organize the framework of our analysis in a way that allows us to be model independent. We 
then discuss our analysis strategy in detail, describing the signal regions and the main backgrounds
involved.

\subsection{Framework}
\label{sec:model}

In order to probe the RPV model, and other new physics scenarios that populate the single-lepton 
channel, it is useful to separate it into two distinct signatures: (a) $1\ell + 1j$ 
({\texttt {SR\_ej}}), and $1\ell + (\geq 3j)$ (\texttt{SR\_e3j}). The branching ratios satisfy:
\begin{align}
    \BR\left(1\ell + 1j\right) + \BR\left(1\ell\; + \geq 3j\right) + \BR\left(\text{other}\right) 
    &= 1\,.
    \label{eq:sum-branching-ratios}
\end{align}
Here, $\BR\!\left(X\right)$ refers to the branching ratio for the resonantly produced squark to 
decay into the final state $X$. $\ell = e$ or $\mu$, and $j$ is any (light) SM jet. Direct decays of the 
squark via an LQD coupling contribute to $\BR(1\ell + 1j)$, just like a leptoquark. From 
\cref{tab:tab1}, we see that the $\BR(1\ell+\!\geq3j)$ channel gives us almost complete coverage 
of the possible cascade modes. $\BR\left(\text{other}\right)$ takes into account 
the squark decays \textit{not} covered by the single-lepton channel, \textit{e.g.}, as in the last 
line of \cref{tab:tab1}.\footnote{An extended RPV sector with multiple non-zero couplings could lead to further squark 
decays, possibly contributing to $\BR\left(\text{other}\right)$. However, note the strict bounds on
products of couplings from flavor changing neutral currents \cite{Allanach:1999ic,Barbier:2004ez} 
and from proton decay~\cite{Chamoun:2020aft,Smirnov:1996bg}.} The separation as in 
\cref{eq:sum-branching-ratios} allows us to experimentally distinguish between pure scalar 
leptoquark theory and a more complicated spectrum and decay pattern as for example in RPV 
superymmetry. 

Analytic expressions exist for the RPV-SUSY 2- or 3-body final states contributing to the branching 
ratios appearing in \cref{eq:sum-branching-ratios} \cite{Richardson:2000nt}. For the more
complicated decay chains, numerical methods are necessary, see for example the tools \texttt{HERWIG} 
\cite{Dreiner:1999qz,Corcella:2000bw}, \texttt{SPheno} \cite{Porod:2003um} and 
\texttt{MadGraph5\_aMC@NLO}~\cite{Alwall:2014hca}. The analytic branching ratios even for the 
simpler decay chains are complicated expressions of the relevant supersymmetric parameters. Thus, we take the branching ratios in \cref{eq:sum-branching-ratios} as our free parameters. This also
underlines our model-independent approach, as these branching ratios could easily be computed in any other model 
leading to the cascade decay of a resonance.

We now discuss the specifics of the search strategy, targeting the two signal regions 
$1\ell + 1j$ ({\texttt{SR\_ej}}) and $1\ell + (\geq 3j)$ (\texttt{SR\_e3j}).

\subsection{Signal Region: \texttt{SR\_ej}}
The $1\ell + 1j$ mode for a squark corresponds exactly to a decaying single leptoquark, as 
investigated in Ref.~\cite{Buonocore:2020erb}. We briefly review this, and implement it 
analogously. We require one negatively charged electron or muon,\footnote{The positively charged
lepton mode is slightly PDF suppressed, by the different luminosities of $u$- and $d$-quarks in 
the proton. Of course, at high energies, one must also consider how well charge identification 
can perform but we consider $100\%$ efficiency here.} and one light jet and label the signature 
as \texttt{SR\_ej}. Since we expect the mass of the squark decaying into the lepton and jet to 
be $\mathcal{O}\left(\SI{1} {\tera\electronvolt}\right)$, we impose rather strict requirements 
on the transverse momentum of both objects: 
\begin{equation}
    p_\text{T}(\ell),\, p_\text{T}^{\mathrm{jet}} > \SI{500}{\giga\electronvolt}\,,
\end{equation} 
with a pseudorapidity cut, $|\eta| <2.5$.
$Z$-boson, top quark, and QCD backgrounds are reduced by imposing a veto on events with an extra
lepton with $p_\text{T} \!>\SI{7}{\giga\electronvolt}$ (and $|\eta| < 2.5)$, or an extra jet 
with $p_\text{T} \!> \SI{30}{\giga\electronvolt}$ (and $|\eta| < 2.5)$. $W$-boson backgrounds are 
reduced by requiring $E^{\text {miss}} _\text{T}<\SI{50}{\giga\electronvolt}$. 

With the above basic cuts, two non-negligible backgrounds remain: Single $W^-$
production in association with jets (with the $W^-$ decaying leptonically), and QCD multijet production, where one of the 
jets is misidentified as a lepton. In Ref.~\cite{Buonocore:2020erb} the other backgrounds 
are plotted; they constitute less than $\mathcal{O}\left(5\%\right)$ of the total background in 
the major part of the phase space. We thus neglect them here. See also the cutflow table corresponding to the benchmark point of~\cref{eq:benchmark} in 
\cref{tab:tab2}.

The strategy for this signal region is to look at the invariant mass distribution
formed by the leading lepton and jet. The signal is expected to peak in a narrow region 
around the squark mass, while the background falls monotonically. We present numerical 
results for our benchmark scenario in \cref{sec:discussion}. 

\subsection{Signal Region: {\texttt{SR\_e3j}}}
In this signal region we require one charged electron or muon (or their antiparticles), and at 
least three jets; we label it as \texttt{SR\_e3j}. Here, we do not restrict ourselves to only the 
negatively charged leptons as the cascades in \cref{tab:tab1} involve Majorana fermions, 
\textit{e.g.}, the neutralinos or the gluino, which decay into a final state or its charge conjugate 
with equal probability.

We implement the following basic cuts for the leading lepton and the three leading jets:
\begin{eqnarray}
p_\text{T}(\ell)&>& \SI{200}{\giga\electronvolt}\,, \\
p_\text{T}^{\mathrm{jet}_1},\,p_\text{T}^{\mathrm{jet}_2},\,p_\text{T}^{\mathrm{jet}_3}
&>&\SI{50}{\giga\electronvolt},
\end{eqnarray} 
with all objects required to have $|\eta| < 2.5$. As before, to reduce $Z$-boson backgrounds, we veto events 
with an extra lepton (satisfying $p_\text{T} \!>\SI{7} {\giga \electronvolt}$ and $|\eta| < 2.5$). 
Top backgrounds are reduced by a $b$-jet veto. However, unlike the \texttt{SR\_ej} case, we do not 
veto events with extra light jets. 

A useful category of cuts is provided by scalar sums of energies of the final state objects.
These mostly depend on the energy scales involved and not on the cascade details. We 
employ two: the sum of $|p_\text{T}|$ of all reconstructed jets, $H_\text{T}$; and 
the total scalar sum of the $|p_\text{T}|$ of all reconstructed objects and the missing transverse energy,
$S_\text{T}$. We require $H_\text{T} \!> \SI{900}{\giga\electronvolt}$, and $S_\text{T} \!>
\SI{1500}{\giga\electronvolt}$. We reduce the $W$-boson background by requiring $E^{\text
{miss}} _\text{T}<\SI{50}{\giga\electronvolt}$.

With the above cuts, to a good approximation, the main backgrounds are associated single $W$, 
associated $t\bar{t}$, and QCD multijet production. One can find the relevant backgrounds 
plotted as a function of $H_\text{T}$ in Ref.~\cite{CMS:2016muu}, where the analysis has some 
overlap with the cuts we make. There, one can see that the other backgrounds make up less than
$\mathcal{O}\left(5\%\right)$ of the total background. Actually, in our case, 
the $t\bar{t}$ background is also expected to be much smaller than the associated $W$ one. In general, the
former only becomes significant relative to the latter, when one requires a large number of jets 
in the final state, or if the jets have lower energies.\footnote{For instance, see the relative 
contributions of the two backgrounds (in the zero $b$-jet tag bin) as a function of the 
number of jets and the energies required in Refs.~\cite{ATLAS:2017oes,ATLAS:2021fbt}.} Since 
our final state of interest only has a few jets and these are energetic, we will focus on 
the $W +$ jets and the QCD multijet backgrounds. See also the cutflow table corresponding to the benchmark point of~\cref{eq:benchmark} in \cref{tab:tab3}; we show the $t\bar{t}$ numbers too for comparison, but we do not include them in our numerical studies.

We note that in designing the above cuts, we have chosen generality over optimality. The kinematical 
configuration of the final state objects is decided by the details of the gauge cascade, which 
in turn depend on the SUSY mass spectrum, to which we choose to remain blind in our approach. Finally, our strategy is to look at invariant mass distributions for the squark and the particle at the end of the cascade (typically the LSP).

\subsection{Other Signatures}
So far we have only focused on resonant squark production at the LHC via the lepton PDFs. Here we briefly mention that a non-zero $\lam'$, as we have been considering, can also lead to resonant \textit{slepton} production via quark PDFs.\footnote{Note that the bound on $\lam'_{111}$ from neutrinoless double beta decay \cite{Hirsch:1995ek,Allanach:1999ic} is strongly model dependent and is not relevant
for a heavy neutralino and gluino as we discuss here.} The 
direct decay gives a resonance bump in the 2-jet cross-section. The cascade decay via a neutralino LSP leads to the promising signature of like-sign dileptons 
\cite{Dreiner:2000vf,Dreiner:2006sv,Dreiner:2012np}. More relevant to the 
search presented here is the decay of the neutralino to a neutrino and 2 jets, giving an overall signature of 1 lepton + 2 jets and $E^{\text{miss}}_\text{T}$ from a neutrino. However, with 2 jets and possibly a large amount of $E^{\text{miss}}
_\text{T}$, this is orthogonal to the search at hand, and we do not further consider 
it. We now present some numerical results.  

%=============================================================================
\section{Results and discussion}
\label{sec:discussion}
	%=============================================================================
\subsection{Numerical Setup}
For the results presented here, we have generated event samples corresponding to center-of-mass energy $\sqrt{s} =\SI{13}{\tera\electronvolt}$, using the program 
\texttt{MadGraph5\_aMC@NLO}~\cite{Alwall:2014hca} linked to 
\texttt{PYTHIA\;8.2}~\cite{Sjostrand:2014zea} for showering and/or decays. Once showered, the 
event samples are passed through our analysis which has been implemented in 
\texttt{CheckMATE\;2}~\cite{Dercks:2016npn,Cacciari:2011ma,Cacciari:2005hq,Cacciari:2008gp,Read:2002hq}; 
detector effects are accounted for by the linked \texttt{DELPHES\;3}~\cite{deFavereau:2013fsa} 
component. For all electrons in our analysis, we have used the \texttt{ATLAS} `Tight' criterion 
implemented in \texttt{CheckMATE\;2} while for jets we use the anti-$k_\text{T}$ algorithm 
implemented in \texttt{FastJet}, which is the \texttt{CheckMATE\;2} default. We choose the cone
size $\Delta R=0.4$.

We have generated the signal samples using the \texttt{UFO} \texttt{RPV-SUSY} model file available 
at Ref.~\cite{FR:RPVMSSM}. We use the lepton PDFs calculated in Ref.~\cite{Buonocore:2020nai}, 
which have been implemented in the \texttt{LHAPDF}~\cite{Buckley:2014ana} setup. One subtle point 
is the handling of initial state leptons during showering; here we have followed the prescription 
described in Ref.~\cite{Buonocore:2020erb}. Note that we have only generated the signal sample at 
leading order (LO). Both NLO (next-to-leading order) QCD and NLO QED 
corrections~\cite{Gauld:2019pgt, Garcia:2020jwr} relying on the photon PDF can be significant, but 
they contribute with opposite signs and comparable magnitudes, leading to a milder-than-expected net correction of $\mathcal{O}\left(10\%\right)$ to the LO cross-section~\cite{Greljo:2020tgv}. This would be important to consider in precision studies; this is not the focus of the present work. 

We define the following benchmark scenario that we use to present most of our results in this
section:
\begin{align}
\lambda'_{111} = 0.4\,,\; M_{\tilde{u}_L} = M_{\tilde{d}_R} = \SI{2}{\tera\electronvolt}
\,,\; M_{\text{LSP}} = \SI{1}{\tera\electronvolt}\,.
\label{eq:benchmark}
\end{align}
The above choice is motivated by current LHC squark limits, while the LSP can be drastically
lighter~\cite{Mitsou:2714089}. The value of $\lam'_{111}$ is chosen to lie near current constraints 
from low-energy experiments~\cite{Hirsch:1995zi,Allanach:1999ic}. We stress that in
the above we do not specify the nature of the LSP, or the details of the SUSY parameters. Instead,
as mentioned earlier, we treat the branching ratios as the free variables that capture all the 
relevant information. We do however assume that no RPV coupling other than $\lam'_{111}$ 
contributes to our two signal regions.
  
For the $W^- +$ jets background corresponding to \texttt{SR\_ej}, we use \texttt{MadGraph5\_aMC@NLO}
to generate one electron, one anti-neutrino plus one jet at LO in QCD. We have not included the electroweak contribution. We deal with the issue of low 
statistics in the high-$p_\text{T}$ region by implementing a generator level cut on the lepton:
$p_\text{T} \!>\SI{400}{\GeV}$, and through phase-space splicing. We split the phase space into 
several regions based on the $p_\text{T}$ of the electron, and then glue them together at the end 
to get a continuous distribution. We account for NLO QCD corrections by employing a $k$-factor. 
In general, $k$-factors are phase-space dependent; to ensure we get a value that is appropriate 
for our region of interest, we use \texttt{MadGraph5\_aMC@NLO} to calculate the total cross-sections 
for the above process at LO and NLO in the phase-space region where the electron has $p_\text{T}
\!> \SI{400}{\GeV}$. Taking the ratio, this gives us a $k$-factor of $1.61$. We have validated our
obtained background against Ref.~\cite{Buonocore:2020erb} and find good agreement. We depict the 
cutflow corresponding to the cuts of \texttt{SR\_ej} for the background and the benchmark 
signal point in \cref{tab:tab2}.

\begin{table}
\begin{center}
\begin{ruledtabular}
\renewcommand{\arraystretch}{1.2}
\begin{tabular}{ccc} 
{\bf Cuts} & {\bf Signal} & {\bf $W^- +$ jets}\\
Generator Level & $91$ & $11050$\\
Leading lepton $p_\text{T} \!> \SI{500}{\giga\electronvolt}$ & $37$ & $3274$\\
Leading jet $p_\text{T} \!> \SI{500}{\giga\electronvolt}$ & $34$ & $2183$\\
$E^{\text{miss}}_\text{T} < \SI{50}{\giga\electronvolt}$ & $21$ & $750$\\
Veto & $10$ & $278$\\
\end{tabular}
\end{ruledtabular}
\end{center}
\caption{Cutflow corresponding to the basic cuts for \texttt{SR\_ej} for $\SI{100}{\femto\barn^{-1}}$ of integrated luminosity. For the signal, the generator level cuts are are the default \texttt{MadGraph5\_aMC@NLO} values. For the associated $W^-$ background, an additional cut of $p_\text{T} \!> \SI{400}{\giga\electronvolt}$ on the lepton has been applied. This is why it appears as if the lepton $p_\text{T}$ cut affects the signal more than the background, degrading the signal to background ratio; this is a superficial effect. Lastly, the final veto step actually dilutes the signal ratio against the $W^-+$ jets BG. However, as explained in the main body, it is crucial in reducing the $Z$-boson, top and QCD backgrounds, which would otherwise dominate over the signal.}
\label{tab:tab2}
\end{table}

For the QCD multijet background,  \texttt{ATLAS} and \texttt{CMS} usually use data-driven
studies over simulation. We use the numbers provided in Ref.~\cite{Buonocore:2020erb}; these 
have been read off from a data-driven study by \texttt{ATLAS} in Ref.~\cite{ATLAS:2013wgh}.

For the $W +$ jets background in the case of \texttt{SR\_e3j}, we use \texttt{MadGraph5\_aMC@NLO} 
linked to \texttt{PYTHIA\;8.2} to generate one electron, one anti-neutrino plus up to $3$ jets (and the charge conjugated process) at LO accuracy in QCD using the MLM
prescription~\cite{Mangano:2001xp,Alwall:2007fs,Alwall:2008qv}, with the xqcut scale set 
to $\SI{70}{\GeV}$; we have checked that this gives smooth differential jet rate (DJR) 
distributions for our process and energy scale~\cite{MLM:Tut}. We have not included the electroweak contribution. To obtain sufficient 
statistics in the tail of the distribution, we again use generator level cuts: lepton $p_
\text{T}\!>\SI{150} {\GeV}$, $iH_\text{T}\!>\SI{800}{\GeV}$, and $E^{\text{miss}}_\text{T}< 
\SI{50}{\GeV}$, and phase-space splicing and gluing---this time relying on splits based on 
the $iH_\text{T}$ (inclusive scalar sum of jet energies) variable available in 
\texttt{MadGraph5\_aMC@NLO}. 

We account again for NLO effects by employing a $k$-factor. Here we take the ratio of the 
total cross-sections for associated $W$ production at NLO and LO with the default 
\texttt{MadGraph5\_aMC@NLO} cuts which gives us a $k$-factor of $1.286$. We expect 
this to be an overestimate since the $k$-factor decreases both in the relevant region of 
phase space~\cite{Huston:2010xp, Berger:2009ep}, as well as when requiring a larger number 
of associated jets. This avoids the computationally intensive task of calculating the full 
NLO cross-section with $3$ extra partons; similar approaches have been employed by 
\texttt{ATLAS} and \texttt{CMS} in Refs.~\cite{ATLAS:2017avu,CMS:2017gbl}, and we emphasize that 
this produces conservative results. We depict the cutflow corresponding to the cuts of
\texttt{SR\_e3j} for the background and the benchmark signal point in \cref{tab:tab3}.
\begin{table}
\begin{center}
\begin{ruledtabular}
\renewcommand{\arraystretch}{1.2}
\begin{tabular}{cccc} 
{\bf Cuts} & {\bf Signal} & {\bf $W +$ jets} & {\bf $t\bar{t}$}\\
Generator Level & $130$ & $9565$ & $2615$\\
$b$ veto & $118$ & $8389$ & $539$\\
Leading lepton $p_\text{T} \!> \SI{200}{\giga\electronvolt}$,\\
Extra lepton veto & $32$ & $3787$ & $114$\\ 
$p^{\text{jet 1,2,3}}_\text{T} \!> \SI{50}{\giga\electronvolt}$ & $29$ & $2562$ & $72$\\
$iH_\text{T} \!>\SI{900}{\giga\electronvolt}$ & $25$ & $1892$ & $26$\\
$S_\text{T} \!>\SI{1500}{\giga\electronvolt}$ & $21$ & $935$ & $10$\\
$E^{\text{miss}}_\text{T} < \SI{50}{\giga\electronvolt}$ & $12$ & $417$ & $3$\\
\end{tabular}
\end{ruledtabular}
\end{center}
\caption{Cutflow corresponding to the basic cuts for \texttt{SR\_e3j} for $\SI{100}{\femto\barn^{-1}}$ of integrated luminosity. For the signal, 
the generator level cuts are the default \texttt{MadGraph5\_aMC@NLO} values. For the 
associated $W$ ($t\bar{t}$) background, extra cuts are used on top of the default ones: Lepton $p_\text{T} \!>
\SI{150}{\giga\electronvolt}$, $iH_\text{T} \!>\SI{800}{\giga\electronvolt}\; (\SI{500}{\giga\electronvolt})$, and $E^{\text 
{miss}}_\text{T} < \SI{50}{\giga\electronvolt}$. Consequently, the detector level cuts appear to suppress the signal more strongly that the background. 
% This is why it superficially seems like the cuts affect the signal more than the backgrounds and degrade the ratio, but this is not the case, of course.
}
\label{tab:tab3}
\end{table}

\begin{figure*}
	\centering
        \includegraphics[width=.495\textwidth]{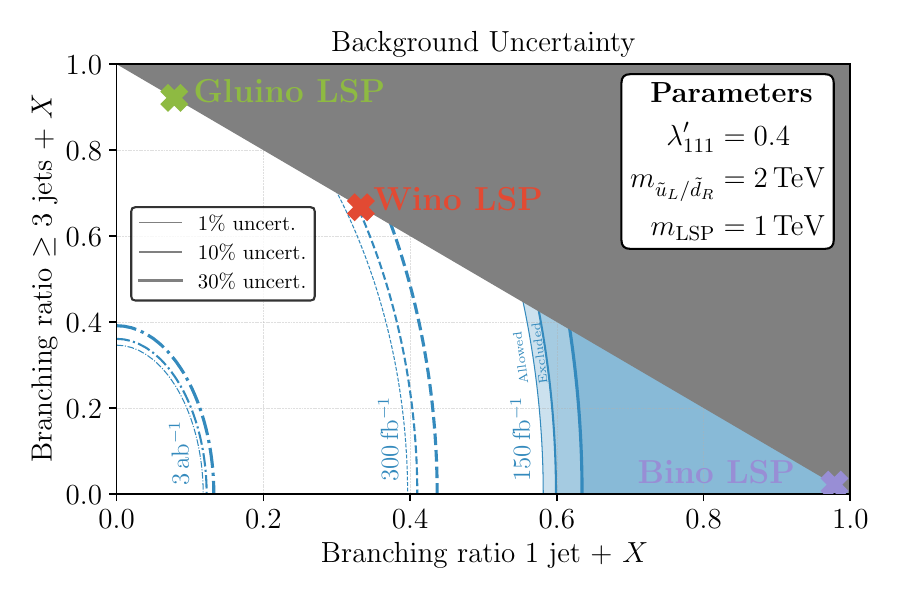}
        \includegraphics[width=.495\textwidth]{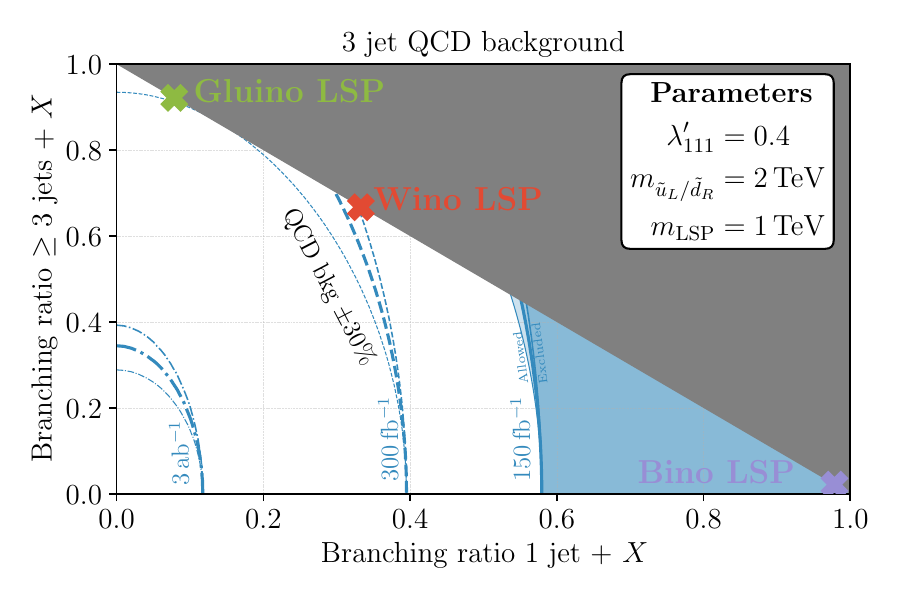}
	\caption{Projected constraints based on integrated luminosities of $(150,300,3000)\SI{}{\,\femto\barn^{-1}}$ shown as blue-$($solid, dashed, dot-dashed$)$ contours. These constraints are shown as a function of the branching ratios into the two channels: ($i$) 1 jet + $X$, ($ii$) $\geq 3$ jets + $X$ where $X$ denotes either an electron or a neutrino. As reference points, we show the expected branching ratios for an RPV model featuring a \SI{1}{\TeV} gluino, bino or wino-LSP, see \cref{sec:model} for more details. \textbf{Left:} We vary the systematic uncertainty on the background predictions in both channels (1\%, 10\% and 30\%) showing how the exclusion limits change. \textbf{Right:} We study how varying the QCD background rate for the three jet channel from the assumed 30\% affects our results. }
	\label{fig:BRlimits}
\end{figure*}

We do not calculate the QCD multijet background but rather include it as an extra $30 \%$
contribution to the final number of background events after our cuts. This is a very rough
estimate using the background distributions plotted in Ref.~\cite{CMS:2016muu}, and 
accounting for the fact that the extra missing energy cut we make in our analysis targets
the $W$-boson more effectively than it will target multijets. We will study the importance of this assumption in~\cref{fig:BRlimits}.

For \texttt{SR\_ej}, after passing the samples through the basic cuts,  we look at the 
invariant mass distribution of the leading lepton and leading jet to reconstruct the 
squark mass. The width of the bins, for a narrow resonance, is determined by the 
experimental resolution. Here, we choose it to be approximately $10\%$ of the invariant 
mass. 

For \texttt{SR\_e3j}, we first reconstruct the invariant mass distribution for the particle at the end of the cascade, choosing a rather broad binning size of $\SI{400}{\giga\electronvolt}$.\footnote{In a spectrum-blind approach, one does not know which final state objects originate from the decay of the cascade-end particle. Experimentally this requires looking at multiple  distributions formed by combinations of the reconstructed electron and jets; see, for instance, Ref.~\cite{CMS:2018wkq}. Here we work with simulations where the cascade-end is fixed which means we do not go through this procedure. However, our chosen broad binning size compensates to account for possible inefficient matching between the reconstructed objects and the true parton level decay products.} For events in each bin, we then reconstruct the squark mass by looking at the invariant mass distribution formed by all reconstructed objects, selecting the binning width to be approximately $10\%$ of the invariant mass.

Finally, we calculate 
the potential exclusion significance~\cite{Cowan:2010js} for both \texttt{SR\_ej} and \texttt{SR\_e3j} by 
reading off the signal and background numbers in each squark bin and select the highest value as the resulting significance. 

\subsection{Search Sensitivity}
We present the projected $95\%$ confidence level (CL) exclusion limits corresponding to 
the benchmark scenario, \cref{eq:benchmark}, for the current data on tape ($\SI{150}{\femto 
\barn^{-1}}$),  as well as projections for the HL-LHC using $\SI{300}{\femto\barn^{-1}}$
and $\SI{3}{\atto\barn^{-1}}$ of integrated luminosity in \cref{fig:BRlimits}.\footnote{This
is assuming no discovery is imminent at the projected reach.}  The projections are shown 
in a model-independent manner as a function of the branching ratios of 
\cref{eq:sum-branching-ratios}. We depict on the figure where a supersymmetric model with 
a $\SI{1}{\TeV}$ gluino-, wino-, or bino-like LSP, respectively, would lie, assuming 
the resonantly produced squark is the NLSP. Here we have assumed that $\lam'_{111}$ is 
the only non-zero RPV coupling, and no other decays are open. Therefore, they all 
lie on the line, 
\begin{equation}\BR\left(X+1j\right)+ 
\BR\left(X+\geq 3j\right) = 1\,.
\end{equation}
We have combined the decay modes corresponding to a charged lepton and a neutrino into a 
single branching ratio, namely $X \equiv \ell$ or $\nu$. This is convenient for the normalization
since in the RPV-MSSM, neglecting lepton masses, the two modes are symmetric to a very good
approximation.
%\footnote{This is because the tree-level decays into the single charged lepton mode occur via the LQD operator which contains an $\mathrm{SU}(2)_{L}$ doublet. At higher-order in the RPV couplings, however, the charged lepton can be right-handed with no neutrino counterpart due to the LLE operators combining with LQD and/or UDD.} 
Our analysis still targets only the charged lepton, \textit{e.g.}, $\BR\left(1\ell+1j\right)=
0.5\times\BR\left(X + 1j\right)$.

We emphasize that \cref{fig:BRlimits} can be used to re-interpret the results for any 
model. For example, for an additional significant non-zero $\lam_{121}$, we would obtain 
decays with more than 1 charged lepton in the final state, resulting in a non-zero BR(other), 
\begin{equation}
\BR(X+1j)+\BR(X+\geq3j)<1\,.
\end{equation}
Such a model would lie in the lower left triangle.

\cref{fig:BRlimits} shows that, even with current data, the single-lepton channel can probe large regions of the RPV model space (for instance, the Bino LSP scenario) corresponding to the benchmark point, allowing us to go beyond existing bounds. By the end of HL-LHC runs, nearly the whole space of models corresponding to the benchmark can be probed. 

The figure also studies how systematic uncertainties, and our assumption about the QCD background in \texttt{SR\_e3j} affect our results.

\begin{figure*}
	\centering
        \includegraphics[width=.495\textwidth]{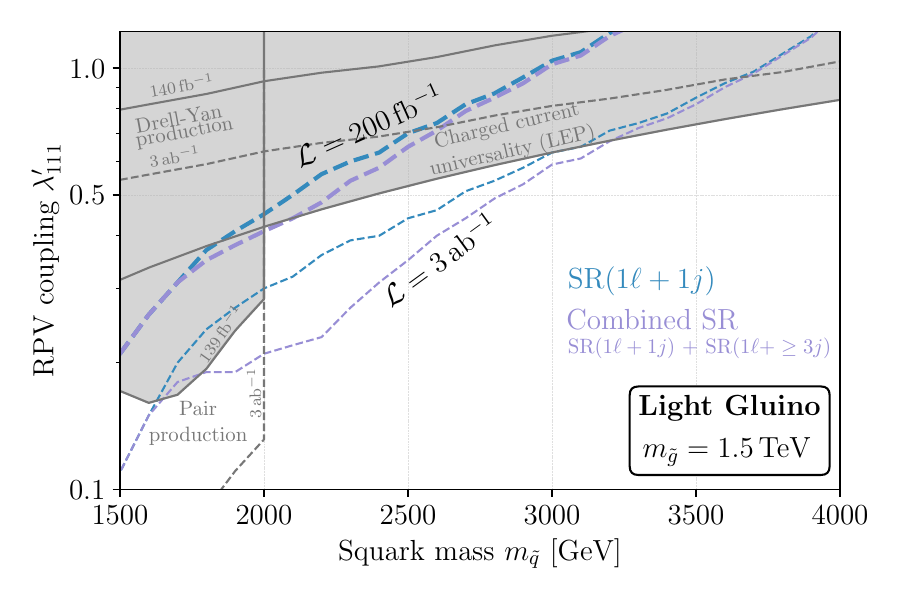}
        \includegraphics[width=.495\textwidth]{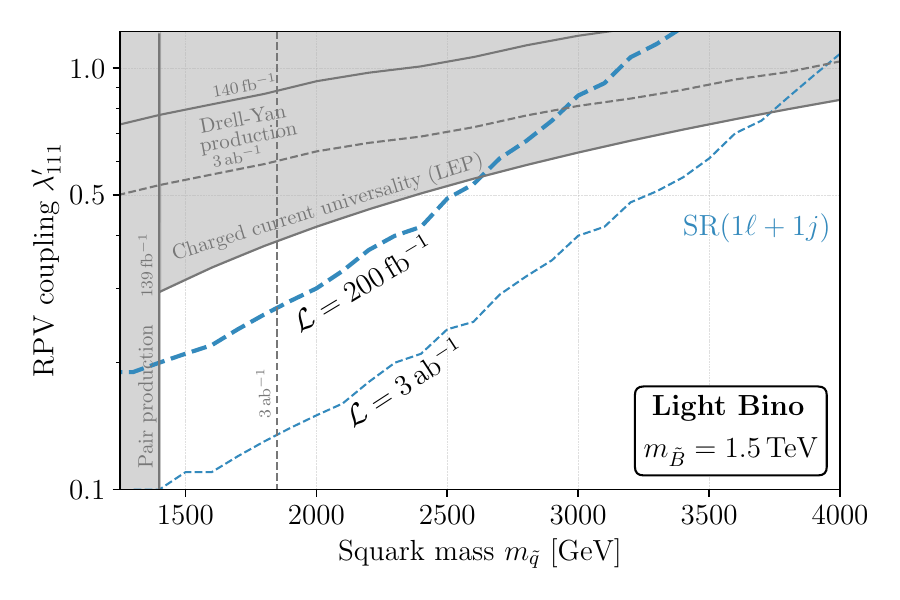}
\caption{The search sensitivities of the single-lepton channel for the light gluino (left) 
and light bino (right) scenarios corresponding to two different integrated luminosities: 
200\,fb$^{-1}$ (thick lines) and 3\,ab$^{-1}$ (thin lines). For the gluino case, the reach of 
the \texttt{SR\_ej} search is shown in turquoise and the combined reach of \texttt{SR\_ej} + 
\texttt{SR\_e3j} is shown in purple. For the bino case we just show the reach of 
\texttt{SR\_ej}. In both plots we show in gray the area excluded by existing experiments: 
Drell-Yan~\cite{CMS:2021ctt}, charged current universality~\cite{Barger:1989rk} and squark
pair production~\cite{ATLAS:2020dsk,Beenakker:2011fu,Bornhauser:2007bf}, with each region labeled by its corresponding cause of exclusion. Finally, we show the projected limits 
assuming $\mathcal{L}=\SI{3}{\atto\barn^{-1}}$ for both Drell-Yan, 
and squark pair production as dashed grey lines.} 
\label{fig:gluinoLSP}
\end{figure*}

In a next step, we study how the exclusion limits depend on the mass of the squarks and the 
RPV coupling. We assume $\tilde{u}_L$ and $\tilde{d}_R$ are mass-degenerate, and consider two simplified setups, corresponding to a 
\SI{1500}{\giga\electronvolt} gluino, and a \SI{1500}{\giga\electronvolt} bino, respectively,
with the rest of the SUSY spectrum decoupled. As before, we also require $\lam'_{111}$ to be 
the only non-zero RPV coupling. The results for the light gluino (left) and light bino (right) 
scenarios are shown in \cref{fig:gluinoLSP}, neglecting systematic uncertainties. We summarize the results of~\cref{fig:gluinoLSP} in~\cref{tab:summary}, depicting the most stringent current bound on $\lambda'_{111}$ and comparing it with the bound implied by the single-lepton search.

\begin{center}
\begin{table*}
  \renewcommand{\arraystretch}{1}
  \begin{adjustbox}{max width=\textwidth}
  \begin{tabular}{c c c c }
    \hline
    \multirow{2}{2cm}{$\mathbf{M_{\tilde{u}_L} = M_{\tilde{d}_R}}$} & \multirow{2}{3.5cm}{\bf $\mathbf{\lambda'_{111}}$ Current Bound} & \multicolumn{2}{c}{\bf $\mathbf{\lambda'_{111}}$ Single Lepton Channel}\\
    % \hline
    % \textbf{Inactive Modes} & \textbf{Description}\\
    % \cline{3-6}
    & & \textbf{\SI{200}{\femto\barn^{-1}}} & \textbf{\SI{3}{\atto\barn^{-1}}}\\
    \hline
    $\SI{1.5}{\tera\electronvolt}$ & 0.17 (0.315) & 0.21 (0.21) & 0.11 (0.11) \\
    $\SI{2}{\tera\electronvolt}$ & 0.28 (0.42) & 0.41 (0.3) & 0.21 (0.15) \\
    $\SI{2.5}{\tera\electronvolt}$ & 0.525 & 0.65 (0.48) & 0.35 (0.24) \\
    $\SI{3}{\tera\electronvolt}$ & 0.63 & 1.0 (0.83) & 0.58 (0.39)\\
    $\SI{3.5}{\tera\electronvolt}$ & 0.735 & 1.49* (1.37*) & 0.8 (0.59)\\\hline
  \end{tabular}
  \end{adjustbox}
  \caption{Summary of~\cref{fig:gluinoLSP}. The table compares the best existing bounds (from pair production, Drell-Yan, and LEP) on $\lambda'_{111}$ and compares it to the bounds implied by the single-lepton search from~\cref{fig:gluinoLSP} for various squark masses, for the light gluino (bino) scenarios. For values marked with an asterisk, the perturbativity constraint is stronger.}
 \label{tab:summary}
\end{table*}
\end{center}

The turquoise line corresponds to the signal region {\texttt{SR\_ej}} while the purple line 
shows the combination of both signal regions {\texttt{SR\_ej} + \texttt{SR\_e3j}}. The thick 
lines show the search sensitivity for an integrated luminosity of $\mathcal{L}=200$\,fb$^{-1}$ 
while the thinner lines correspond to $\mathcal{L}=3$ ab$^{-1}$. The figure also depicts other
current relevant bounds as shaded gray regions. Currently the most stringent constraints at 
large squark masses come from charged current universality measurements at LEP 
\cite{Barger:1989rk}. We have also recast limits from existing pair production~\cite{ATLAS:2020dsk} 
and Drell-Yan searches~\cite{CMS:2021ctt}, as well as a projection of their reach at HL-LHC determined by assuming 
$\mathcal{L}=\SI{3}{\atto\barn^{-1}}$. 

For the light gluino case, the most constraining current exclusion limits are from charged 
current universality, reaching values of $\lam'_{111}>0.3$ for $m_{\tilde{q}}\sim\SI{1500} 
{\giga\electronvolt}$ and $\lam'_{111}>0.8$ for $m_{\tilde{q}}\sim \SI{4000}{\giga
\electronvolt}$, and from squark pair production that is powerful for low masses, reaching 
$\lam'_{111}>0.16$ for masses of the squarks between $1600-1700\,\SI{}{\giga\electronvolt}$.
The pair-production exclusion region below about $\SI{2000}{\giga\electronvolt}$ has a slope because the search relies on the direct decay mode of the squark; higher mass squarks need a higher RPV coupling to have a sufficient branching ratio for this mode. For squark 
masses between $1500-1600\,\SI{}{\giga\electronvolt}$, the exclusion deteriorates slightly. 
The experimentally observed data in this regime are above the expected ones while for higher 
masses, both observed and expected match; see Ref.~\cite{ATLAS:2020dsk}. The search only excludes 
squark masses up to $\SI{2000}{\giga\electronvolt}$. Thus, the shaded area extends vertically 
at this point. 

We see that the single-lepton channel probes phase-space regions complementary
to those probed by pair production and Drell-Yan, as explained in
\cref{sec:single-lepton-channel}. Further, it has the potential to compete with/outdo the
existing charged current universality constraints, with the added advantage of being 
a direct search. 

For a light gluino, the squarks can have a significant branching ratio into the cascade 
mode since it proceeds via the strong coupling. To see this, we depict the sensitivity 
contour corresponding to \texttt{SR\_ej} alone (turquoise), and to \texttt{SR\_ej} + 
\texttt{SR\_e3j} combined (purple). For low squark masses, \texttt{SR\_e3j} is relatively 
unimportant since the squarks have no phase space to decay into the gluino; the direct 
decay dominates. As the mass increases, the relative importance of \texttt{SR\_e3j} grows.
At very high masses, it becomes less important again because the large $\lam'_{111}$ coupling -- required to have a sufficiently high signal rate -- leads to the direct decay rate increasing as well.

The light bino case is shown in the right plot of Fig.~\ref{fig:gluinoLSP}. Current searches 
such as Drell-Yan and charged current universality behave as in the gluino LSP case and so they 
cover almost the same parameter space. However, the squark pair production only reaches values up to $\SI{1400}{\giga\electronvolt}$. As the gluino is now decoupled, the contribution due to $t$-channel gluino exchange is missing leading to a smaller cross-section. Further, there is no dependence on the coupling since, with the bino kinematically inaccessible in this region, the squark dominantly 
decays directly with branching ratio nearly $1$. 

As before, the single-lepton channel is complementary to the existing searches, extending the potential reach. However, the only mode with power of exclusion here is \texttt{SR\_ej}. The branching ratio of the cascade mode via the bino 
is small leading to a low sensitivity of \texttt{SR\_e3j}. Correspondingly, we have not included the \texttt{SR\_e3j} curves. In comparing with the gluino LSP case on the 
left, we see that the single-lepton channel excludes more parameter space here. This is 
because of the higher branching ratio of the direct decay, contributing to \texttt{SR\_ej}; this mode has a cleaner signature and hence higher exclusion potential than \texttt{SR\_e3j}. 

We note that our proposed search outperforms high-luminosity 
projections of the searches based on Drell-Yan and squark pair production (see dashed gray 
lines in \cref{fig:gluinoLSP}). The Drell-Yan constraints begin to be competitive at very high masses; although they are still surpassed by constraints from LEP measurements. On the other hand, for pair production, the reach improves for low squark masses, reaching $m_{\tilde{q}} 
=\SI{2}{\TeV}$ in the light gluino case, and $m_{\tilde{q}}=\SI{1.8}{\TeV}$ in the light 
bino case. These projections are based on current searches, which place a strict cut of 
\SI{2}{\TeV} on the squark masses. This limits the sensitivity of our projections. 
Nevertheless, this last search is powerful for low masses exhibiting strong complementarity 
with the search proposed in this work. 

An interesting observation is that \texttt{SR\_ej} is quite powerful even in cases where the squark 
has low BR into the direct mode, \textit{e.g}, the light gluino scenario, left plot in
Fig.~\ref{fig:gluinoLSP}. This shows that a simple resonance $1 \ell + 1j$ search is 
also a powerful probe of the entire RPV space, even with a more complicated spectrum; not 
just the leptoquark-like scenario. On the other hand, \texttt{SR\_e3j} apart from extending 
the reach of the searches (particularly at high-luminosities), will be crucial to distinguish between 
leptoquarks and squarks in case of a discovery.

It is also important to emphasize that even though we have considered simplified setups, with most of the SUSY spectrum decoupled, 
our results are more general. For instance, we would realistically expect the sleptons and
electroweakinos to also be light in the light gluino case. In such a scenario, new gauge-cascade
chains can open up for the resonant squark, thus diluting the direct and gluino decay modes.
However, the signals from these distinct chains will simply add up with those from the gluino 
mode in the \texttt{SR\_e3j} bin, as long as the end point of all the cascades is the same. 
This is exactly what happens, for instance, for small RPV couplings, where all gauge-cascade 
chains end in the LSP.

Finally, we note that one can perform completely analogous studies for LQD operators involving 
second generation fermions. For a coupling with a second generation lepton, \textit{e.g.}, $\lam'_{211}$ we expect the limits to weaken only slightly as the muon PDFs are only mildly suppressed relative to the electron ones \cite{Buonocore:2020nai}. The case of second generation quarks is slightly more involved. For $\lam'_{111}$ the dominant production mode is $e u \rightarrow \tilde{d}$ versus $e 
d \rightarrow \tilde{u}$, roughly in the ratio 2:1. Thus, we would expect the case $\lam'_{112}$ to be only mildly suppressed compared to our present analysis, whereas 
the case $\lam'_{121}$ to be more suppressed. But both should still be feasible; see Ref.~\cite{Buonocore:2020erb} for quantitative estimates.

\section{Conclusions}
\label{sec:conclusions}

In this paper we have shown that the single-lepton channel is a promising signature in the 
search for new physics beyond the Standard Model. We have considered two specific versions 
of this channel: (a) A single first or second generation charged lepton, exactly 1 jet and 
low missing transverse energy, which we denoted \texttt{SR\_ej}, and (b) A single first or 
second generation charged lepton, at least 3 jets, and low missing transverse energy, 
\texttt{SR\_e3j}. Utilizing the lepton parton distribution functions (PDF) of the proton, we 
showed that the channel \texttt{SR\_ej} is promising not only in the search for a single 
leptoquark or a directly decaying squark, but remains sensitive even when more complicated supersymmetric cascade decays are accessible. Further, the channel \texttt{SR\_e3j} plays an important role in increasing both the reach and coverage in such scenarios. More importantly, it also acts as a discriminant between a bare scalar leptoquark theory versus one with a more extensive supersymmetric sector featuring
kinematically accessible particles beyond just a light squark. 
 
Although \texttt{ATLAS} and \texttt{CMS} have performed single-lepton searches associated with
large jet multiplicity, see Refs.~\cite{ATLAS:2017oes,CMS:2017szl,ATLAS:2021fbt,CMS:2021knz}, our proposed search covers a variety of scenarios which would not produce a sufficiently 
large number of jets. Beyond the question of coverage, the resonant $s$-channel production mechanism 
invoked in our analysis benefits tremendously from the forthcoming increase of luminosity at the LHC. 
We therefore strongly advocate that this type of search be pursued at forthcoming LHC runs, 
as well as emphasize the necessity of more exhaustive theoretical work surveying the opportunities that will arise in the era of High-Luminosity LHC.

\section*{ACKNOWLEDGMENTS}
We thank Giulia Zanderighi for very helpful comments on their paper~\cite{Buonocore:2020erb} 
and on the numerical implementation of the lepton PDFs, Rhorry Gauld for 
helpful comments about the leptonic PDFs of the proton, both Philip Bechtle and Admir Greljo for 
useful discussions about search strategies and other relevant limits, and lastly Simon Knapen for comments on the manuscript. The work of HKD and SN has
been supported by BMBF project grant 05H18PDCA1. VML is funded by the Deutsche
Forschungsgemeinschaft (DFG, German Research Foundation) under Germany’s Excellence Strategy - EXC 2121 “Quantum Universe” - 39083330. VML also acknowledges the financial support by
Ministerio de Universidades and “European Union-
NextGenerationEU/PRTR” under the Grant No. María
Zambrano ZA2021-081, and the Spanish Grants
No. PID2020–113775GB-I00 (AEI) and No. CIPROM/
2021/054 (Generalitat Valenciana).

%-----------------------------------------------------------------------------
\bibliographystyle{JHEP}
\bibliography{refs}
%-----------------------------------------------------------------------------

\end{document}